\documentclass[pra, showpacs,twocolumn, superscriptaddress,groupedaddress, noshowpacs]{revtex4}  % for review and submission

\usepackage{amsmath,amssymb,amsfonts,amsthm}
\usepackage{braket}
\usepackage{cancel}
\usepackage{color}
\usepackage{graphicx}
\usepackage{hyperref}

\usepackage{physics}

\newcommand {\fabs}[1] {\left| #1 \right|}

\newcommand {\fabsq}[1] {\left\vert #1 \right\vert^2}

\newcommand{\braXket}[3]{\langle#1|#2|#3\rangle}

\begin{document}

\title{Spin squeezing in internal bosonic Josephson junctions via enhanced shortcuts to adiabaticity}

\author{Manuel Odelli}
\affiliation{School of Physics, University College Cork, Cork, T12 K8AF, Ireland}

\author{Vladimir M. Stojanovi\'c}
\email{vladimir.stojanovic@physik.tu-darmstadt.de}
\address{Institut f\"{u}r Angewandte Physik, Technical University of Darmstadt,
64289 Darmstadt, Germany}

\author{Andreas Ruschhaupt}
\address{School of Physics, University College Cork, Cork, T12 K8AF, Ireland}

\begin{abstract}
We investigate a time-efficient and robust preparation of spin-squeezed states -- a class
of states of interest for quantum-enhanced metrology -- in internal bosonic Josephson junctions
with a time-dependent nonlinear coupling strength between atoms in two different hyperfine states.
We treat this state-preparation problem, which had previously been addressed using shortcuts to
adiabaticity (STA), using the recently proposed analytical modification of this class of quantum-control
protocols that became known as the enhanced STA (eSTA) method. We characterize the state-preparation 
process by evaluating the time dependence of the coherent spin-squeezing and number-squeezing 
parameters and the target-state fidelity. We show that the state-preparation times obtained using 
the eSTA method compare favourably to those found in previously proposed approaches. We also 
demonstrate that the increased robustness of the eSTA approach -- compared to its STA
counterpart -- leads to additional advantages for potential experimental realizations of strongly
spin-squeezed states in bosonic Josephson junctions.
\end{abstract}

\maketitle

\section{Introduction}
Time-efficient and robust protocols for quantum-state engineering~\cite{Li+Song:15,StojanovicPRL:20,Peng+:21,StojanovicPRA:21,
Zheng++:22,Haase++:22,Zhang+:23,Zhang++:23,Erhard+:18,Macri+:18,Haase+:21,Qiao+:22,Feng+:22,Nauth+Stojanovic:22,Stojanovic+Nauth:22,
Wu+:17,Wang+Terhal:21,Stojanovic+Nauth:23} represent an important building block of emerging quantum technologies for computing,
sensing, and communication~\cite{Dowling+Milburn:03}. For instance, motivated by their potential quantum-technology applications
a multitude of robust schemes for the preparation of highly entangled multiqubit states of $W$~\cite{Li+Song:15,StojanovicPRL:20,
	Peng+:21,StojanovicPRA:21,Zheng++:22,Haase++:22,Zhang+:23,Zhang++:23}, Greenberger-Horne-Zeilinger (GHZ)~\cite{Erhard+:18,Macri+:18,
	Haase+:21,Qiao+:22,Feng+:22,Nauth+Stojanovic:22,Stojanovic+Nauth:22}, or Dicke~\cite{Wu+:17,Wang+Terhal:21,Stojanovic+Nauth:23}
type have been proposed in recent years; these schemes are of relevance for diverse physical platforms for quantum information
processing. Another interesting class of quantum states are, e.g., large Schr\"{o}dinger cat states, whose realization was already
demonstrated in the past with microwave photons~\cite{Vlastakis+:13} and with mechanical degrees of freedom~\cite{Leong+:20}.

% ------- Spin-squeezed states -------------

Spin-squeezed states~\cite{Gross:12} were originally introduced by Kitagawa and Ueda~\cite{Kitagawa+Ueda:93} as a means of
redistributing the fluctuations of two orthogonal spin directions between each other. Shortly afterwards it was demonstrated
that they enable an enhancement in the precision of atom interferometers that measure time, distance, or magnetic
fields~\cite{Wineland+:92,Wineland+:94} -- the same feature that had been discovered for photonic squeezed states in optical interferometry
a decade earlier~\cite{Caves:81}. More precisely, they provide phase sensitivities beyond the so-called standard quantum limit
$\Delta\theta_{\textrm{SQL}}=1/\sqrt{N}$, which is characteristic of probes that involve a finite number $N$ of uncorrelated
or classically correlated particles~\cite{Giovanetti+Lloyd+Maccone:06}. The realization that spin-squeezed states allow the
possibility to overcome this classical bound, which is inherent to current two-mode atomic sensors~\cite{Szigeti+:21}, established
these states as a useful resource for quantum metrology~\cite{Pezze+:18}. Finally, the intimate connection between spin squeezing
and entanglement -- the ingredient also shared by other types of states that defy the standard quantum limit (e.g. GHZ
states)~\cite{Lee+Kok+Dowling:02} -- was subsequently unravelled~\cite{Sorensen+:01,Sorensen+Moelmer:01}. More recent studies
of spin squeezing have unravelled interesting connections with seemingly unrelated concepts, such as spontaneous breaking of
a continuous symmetry~\cite{Comparin+:22} or quantum scrambling~\cite{Li++:23}.

The tunability of elastic atom-atom collisions in Bose-Einstein condensates (BECs) allows one to generate a metrologically
useful entanglement in these systems~\cite{Pezze+:18,Fadel+:18,zhang2023}. In particular, the generation of spin-squeezed 
states was demonstrated in groundbreaking experiments with cold neutral atoms in BECs more than a decade 
ago~\cite{Esteve+:08,Gross+:10,Riedel+:10,Zibold+:10}. These experimental demonstrations were performed with interacting 
cold $^{87}$Rb atoms in bosonic Josephson junctions (BJJs)~\cite{Gati+Oberthaler:07},
a class of systems where bosons within a condensate can be restricted to occupy only two single-atom states (modes). Importantly,
these proof-of-principle experiments were carried out both with internal BJJs [those in which two modes correspond to two different
internal (hyperfine) atomic states and are linearly coupled] and with external ones (where bosons are trapped in two spatially
separated wells of an external double-well potential). More recent experiments with these systems demonstrated more strongly 
spin-squeezed states than the pioneering ones, in addition to reducing the levels of noise from various sources~\cite{Ockeloen+:13,Muessel+:15}.

% ----- Review eSTA ---------------

A specific class of techniques to design control schemes are Shortcuts to Adiabaticity (STA)~\cite{Chen2010STA,
chen2010b,ibanez2012,torrontegui2013a} (for a review, see Ref.~\cite{STA_RMP:19}).
These are analytical control techniques that mimic adiabatic evolution on much shorter timescales. Analytical control solutions are particularly
desirable as they are simpler, provide greater physical insight and allow for additional stability requirements \cite{ruschhaupt2012a,lu2020}.
STA have been applied in many different contexts, e.g. \cite{li2022,kiely2016,kiely2018,torrontegui2011}. In particular, STA-based proposals
for the generation of spin-squeezed
states in BJJs were put forward~\cite{JuliaDiaz+:12,JuliaDiaz++:12}, along with proposals for engineering other types of entangled states
(such as, e.g., NOON states~\cite{Hatomura:18,Stefanatos+Paspalakis:18}), leading to results comparable to those obtained using optimal
control~\cite{Lapert+Ferrini+Sugny:12,Sorelli+:19}. However, experimental realizations of these STA-based proposals for quantum-state
engineering in BJJs have not been reported to date.

However, STA methods can have limitations as they could require non-trivial physical implementation (e.g. counterdiabatic driving \cite{STA_RMP:19}),
while other STA techniques may only be easily applied to small or highly symmetrical systems (e.g. Lewis-Riesenfeld invariants) \cite{STA_RMP:19}.
This motivated the development of Enhanced Shortcuts to Adiabaticity (eSTA) \cite{Whitty+:20, Whitty+:22, Whitty2022b} which allows to correct
STA solutions perturbatively and in an analytical way, inspired by optimal-control techniques~\cite{Werschnik+Gross:07}. The primary motivation
behind this new method is the desire to design efficient control protocols for systems to which STA protocols are not directly applicable.
More specifically yet, the main idea of eSTA is to first approximate the original Hamiltonian of such a system by a
simpler one, tractable within the STA framework. Assuming that the STA protocol for the approximate Hamiltonian is 
close-to-optimal even when utilized for the treatment of the full system Hamiltonian, the actual eSTA protocol is 
obtained through a gradient expansion in the control-parameter space. The eSTA method has already proven superior 
to its STA counterpart in several realistic quantum-control problems pertaining to coherent atom transport in optical
lattices~\cite{Whitty+:20,Hauck+:21,Hauck+Stojanovic:22} and anharmonic trap expansion~\cite{Whitty2022b}. In addition,
it was demonstrated that this approach is intrinsically more robust to various sources of imperfections than its STA
counterpart~\cite{Whitty+:22}.

In this paper, motivated by the dearth of experimental realizations of STA-based quantum-state-engineering schemes in
BJJs, we revisit the problem of efficiently generating spin-squeezed states in internal BJJs. In order to design
a state-preparation scheme that is more amenable to experimental realizations we apply the formalism of eSTA to provide
a control scheme with an improved fidelity and an increased robustness to unavoidable imperfections in current 
cold-atom systems. By making use of the eSTA approach, we determine the time dependence of the interspecies atom-atom 
interaction strength that allows one to time-efficiently generate spin-squeezed states in an internal BJJ. We quantify
the state-preparation process by evaluating the time dependence of the target-state fidelity, as well as that of the 
coherent spin-squeezing and number-squeezing parameters. In this manner, we show that spin-squeezed states can be engineered
in BJJs within times significantly shorter than those in the existing experimental realizations of such 
states~\cite{Esteve+:08,Gross+:10,Riedel+:10}. We also demonstrate that our eSTA-based control scheme yields 
better results than its STA counterpart, both in terms of achievable squeezing and robustness to various types of 
systematic errors, most prominently a systematic control-amplitude error.

The remainder of the present paper is organized as follows.  In Sec.~\ref{SystemProblem}, after briefly reviewing the physics
of an internal BJJ, we introduce its governing Bose-Hubbard-type Hamiltonian, recasting in the Lipkin-Meshkov-Glick form. We then
invoke the well-known mapping of the two-site (or two-state) Bose-Hubbard model of a BJJ to a fictitious particle in a nearly
harmonic potential, which is described by a single-particle Schr\"{o}dinger-like equation in Fock space. Finally, we also introduce
some figures of merit for characterizing spin squeezing. In Sec.~\ref{ApplicationESTA} we apply the eSTA formalism to the system
under consideration. The squeezed-state fidelities computed within the proposed eSTA-based control scheme are then presented and
discussed in Sec.~\ref{sect_results}, where we also demonstrate the extraordinary robustness of our proposed,
eSTA state-preparation scheme. We summarize the paper, with some concluding remarks and outlook, in
Sec.~\ref{SummConcl}.

\section{System and spin-squeezed states} \label{SystemProblem}
\subsection{Internal BJJs and their underlying many-body Hamiltonian} \label{IntroIBJJ}
\begin{figure}[b!]
\centering
\includegraphics[width=0.6\linewidth]{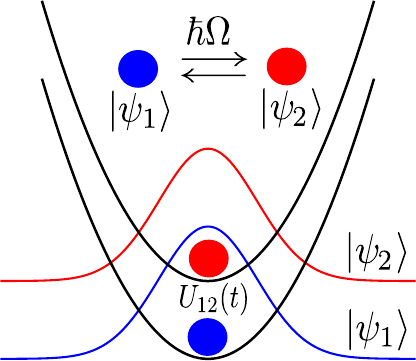}
\caption{\label{fig:sketch} Pictorial illustration of an internal BJJ: An external electromagnetic field coherently 
transfers atoms between the two relevant hyperfine states (modes), denoted by $|\psi_1\rangle$ and $|\psi_2\rangle$,
via Rabi rotations; $\hbar\Omega$ is the corresponding constant Rabi-coupling strength. The interspecies interaction 
strength $U_{12}(t)$, which is assumed to be time-dependent, can be varied using an external magnetic field in the 
presence of a Feshbach resonance. } 
\end{figure}

Internal BJJs are created with trapped BECs in two different internal (hyperfine) states~\cite{Steel+Collett:98}, 
e.g. a condensate of $^{87}$Rb atoms in the $|F=1, m_{\textrm F}=1\rangle$ and $|F=2, m_{\textrm F}
=-1\rangle$ hyperfine levels of the electronic ground state of rubidium. Such a system is typically prepared in practice 
by trapping the atoms in the two internal states in the 
wells of a deep one-dimensional optical lattice; the lattice depth should be sufficiently large that there is no tunnelling 
coupling between different wells~\cite{Gross+:10}. The Josephson-like coupling in those systems is enabled by an electromagnetic 
field that coherently transfers particles between the two relevant internal states -- to be denoted in what follows by 
$|\psi_1\rangle$ and $|\psi_2\rangle$ -- via Rabi rotations~\cite{Hall+:98}. Under the assumption that the external 
motion of the atoms in such a system is not influenced by internal dynamics, it is permissible to use a single-mode 
approximation for each atomic species (i.e., each hyperfine state). A pictorial illustration of an internal BJJ
is provided in Fig.~\ref{fig:sketch}.

% ---- Sketch plot of the two hyperfine states in the internal BJJ ----

The many-body Hamiltonian of an internal BJJ is given by that of a two-state Bose-Hubbard model, which 
using the Schwinger-boson formalism can be recast in the form that represents a special case of the 
Lipkin-Meshkov-Glick family of Hamiltonians~\cite{Lipkin+Meshkov+Glick:65}:
\begin{equation}\label{eq:HamiltonianJosephsonJunctionSTA}
H_{\textrm{IBJJ}}(t) =  U(t) J_z^2 - \hbar \Omega J_x \:.
\end{equation}
Here $\hbar\Omega$ is the (constant) strength of Josephson-like (Rabi) coupling and $U(t)$ is the nonlinear coupling
(i.e. two-body interaction) strength, which we assume to be time-dependent in what follows (we also assume repulsive
interactions, hence $U(t)>0$, at initial and final time); $\mathbf{J}\equiv\{J_x,J_y,J_z\}$ 
are pseudoangular-momentum (collective spin) operators, given in terms of the boson creation and annihilation operators 
corresponding to the two modes ($a^{\dagger}_i$ and $a_i$, respectively, where $i=1,2$) as
\begin{eqnarray} 
J_x &=& \frac{1}{2}\:(a^{\dagger}_1 a_2 +  a^{\dagger}_2 a_1 ) \:, \nonumber \\
J_y &=& \frac{1}{2i}\:(a^{\dagger}_1 a_2 - a^{\dagger}_2 a_1 ) \:, \label{PseudoAngular}\\
J_z &=& \frac{1}{2}\:(a^{\dagger}_1 a_1 - a^{\dagger}_2 a_2 ) \:. \nonumber
\end{eqnarray}
Thus, the Hamiltonian of the system at hand is given by the sum of the one-axis twisting term $U(t)J_z^2$ 
and the Josephson (Rabi-coupling) term $- \hbar\Omega J_x$. The dimensionless 
parameter $\Lambda \equiv NU/(\hbar\Omega)$ quantifies the relative importance of interactions and Rabi 
coupling in an internal BJJ~\cite{Gross:12}. In particular, in the problem at hand where $U=U(t)$ is 
assumed to depend on time, the parameter $\Lambda$ will also be time-dependent. 

The Rabi-coupling parameter $\hbar\Omega$ and interaction strength $U$ in the Hamiltonian
of Eq.~\eqref{eq:HamiltonianJosephsonJunctionSTA} are given by
\begin{eqnarray}
\Omega &=&  \Omega_R \int d^{3}\mathbf{r}\:\psi^{*}_1(\mathbf{r})\psi_2(\mathbf{r}) \:, \nonumber \\
U &=& U_{11}+ U_{22} - 2 U_{12}\:, \label{BJJparameters}\\
U_{ij} &=& \frac{2\pi\hbar^2 a^{(i,j)}_s}{M} \int d^{3}\mathbf{r}|\psi_i(\mathbf{r})|^2
|\psi_j(\mathbf{r})|^2 \:, \nonumber
\end{eqnarray}
where $\Omega_R$ is the Rabi frequency, $a^{(i,j)}_s$ ($i,j = 1, 2$) are intraspecies ($i=j$) and
interspecies ($i\neq j$) $s$-wave scattering lengths, and $\psi_{1,2}(\mathbf{r})\equiv \langle
\mathbf{r}|\psi_{1,2}\rangle$ are mode functions of the two internal states. In particular, the
interspecies $s$-wave scattering length $a^{(1,2)}_s$ for $^{87}$Rb atoms in internal BJJs can be varied using
an external magnetic field owing to the presence of Feshbach resonance \cite{Zibold+:10}; 
in practice, this mechanism is used to reduce the interspecies $s$-wave scattering length, because 
for $^{87}$Rb atoms there is a nearly perfect compensation of intraspecies and interspecies interactions. 
In this manner one can externally control the time dependence $U(t)$ 
of the nonlinear coupling parameter [cf. Eq.~\eqref{eq:HamiltonianJosephsonJunctionSTA}]. An 
alternative approach for tuning the nonlinear coupling strength entails controlling the wave-function overlap 
between the two relevant internal states in a state-dependent microwave potential~\cite{Riedel+:10}; this 
approach also works in magnetic traps and for pairs of internal atomic states for which no convenient Feshbach 
resonances exist.

The control task that we will be concerned with in the remainder of this work amounts to designing 
a control function $U(t)$ [or, equivalently, the time dependence of the dimensionless parameter 
$\Lambda(t)$] such that we start at $t=0$ in the ground state of the Hamiltonian $H_{\textrm{IBJJ}}$ 
[cf. Eq.~\eqref{eq:HamiltonianJosephsonJunctionSTA}] with a given $U(t=0)\equiv U_0=0$ (coherent spin 
state with equal atomic populations in the two relevant internal states) and end up at $t=t_f$ in the 
ground state of that Hamiltonian with a different given $U(t=t_f)\equiv U_f$ (the desired 
spin-squeezed state). While this is not generically the case in control problems treated using the STA- 
and eSTA methods~\cite{STA_RMP:19}, both the initial and final states in our envisioned scheme for the
preparation of spin-squeezed states correspond to ground states of the Hamiltonian of the system. 
This obviates the need for stopping or freezing the system dynamics once the desired spin-squeezed 
state is generated, a feature that makes our state-preparation scheme more flexible and robust.

\subsection{Mapping to a Schr\"{o}dinger-like equation in Fock space}
In the following, we will review the well-known mapping of the two-site (or two-state)
Bose-Hubbard model of a BJJ to a fictitious particle in a nearly harmonic potential, which
is described by a single-particle Schr\"{o}dinger-like equation in Fock space~\cite{Mahmud+:05,Shchesnovich+Trippenbach:08}.

The time-dependent Schr{\"o}dinger equation of this system is written as
\begin{eqnarray}
i \hbar \frac{\partial}{\partial t} \ket{\Psi} = H_{\textrm{IBJJ}}\ket{\Psi} \:.
\label{eq:SchrodingerEquationBoseHubbardAngular}
\end{eqnarray}
By making use of the pseudoangular momentum approach [cf. Sec.~\ref{IntroIBJJ}], a system of $N$ particles can be 
described as a single particle with spin $ N/2 $ and the basis set is of the form $ \{\ket{m}\} $, with $ m = -N/2, ..., N/2 $ 
being the eigenvalues of the operator $J_{z}$. The general state $ \ket{\Psi}$ can be written as
\begin{equation}\label{eq:GeneralStateAngularMomentum}
\ket{\Psi} = \sum_{m = -N/2}^{N/2} c_{m}\ket{m} \:,
\end{equation}
with $\sum_{m = -N/2}^{N/2} \left|c_{m}\right|^2 = 1$ and $\braket{m}{m'} = \delta_{m,m'}$.
By making use of Eqs.~\eqref{eq:SchrodingerEquationBoseHubbardAngular} and \eqref{eq:GeneralStateAngularMomentum},
we readily obtain
\begin{eqnarray}
	\lefteqn{i \hbar \frac{d}{dt}c_{m}(t)} && \nonumber\\
	&=&  - \frac{\hbar \Omega}{2} \left[ \beta_{m} c_{m+1}(t) + \beta_{m-1}c_{m-1}(t) \right]
	+ U(t) m^2 c_{m}(t) \:,\nonumber\\
	\label{eq:numerical}
\end{eqnarray}
where
\begin{eqnarray}
	\beta_m &=& \langle m|J_{-} |m + 1\rangle = \langle m+1|J_{+} |m\rangle \nonumber\\
	&=& \sqrt{\left(\frac{N}{2} + m +1\right)	\left(\frac{N}{2} -m\right)}.
	\label{eq:LadderOperatorsProjected}
\end{eqnarray}
This result yields a Schr{\"o}dinger equation for the coefficients $ c_{m}(t)$, where $m=-N/2, -N/2+1,...N/2$.
Note that Eq.~\eqref{eq:numerical} [resp. the equivalent Eq.~\eqref{eq:SchrodingerEquationBoseHubbardAngular}]
will also be used for the numerically exact simulation of the system at hand in this work.

We now proceed to rewrite this discrete formulation of Eq.~\eqref{eq:numerical} as a continuous one
by performing a change of variables. To this end, we define $h=\frac{1}{N/2}$ and $z_m = \frac{m}{N/2}$ 
(the relative population difference between the two relevant hyperfine states); we then 
find that $z_{m\pm 1} = z_m \pm h$ and $-1 \le z_m \le 1$. By also defining
\begin{eqnarray}
b_{h} (z_m) = \frac{\beta_m}{N} =
\frac{1}{2}\sqrt{\left(1 + z_m +h\right)
\left(1 -z_m\right)} \:, \label{eq:CoefficientCollectingN}
\end{eqnarray}
we obtain that $\beta_{m-1}/N = \frac{1}{2} \sqrt{ \left(1 + z_m \right)\left(1 - z_m -h\right) } = b_{h}(z_m-h)$.
Additionally, we introduce $\psi( t, z_m) = \sqrt{N/2}\:c_{m}( t)$ and straightforwardly verify that $\psi( t, z_m \pm h)= 
\sqrt{N/2}\:c_{m\pm1}( t)$. In addition, we have that $\sum_{m=-N/2}^{N/2} h \left|\psi( t,z_m)\right|^2 = 1$, 
because $\sum_{m=-N/2}^{N/2} \left|c_m (t)\right|^2 = 1$. 

At this point we can switch to the continuum approximation for the relative 
population difference ($z_m\rightarrow z$), introducing at the same time a dimensionless time 
$\tau$, such that $t= 2\tau / \Omega$. In this manner, we can finally recast Eq.~\eqref{eq:numerical} 
in the form
\begin{eqnarray}
\lefteqn{i h\partial_\tau \psi(\tau,z)} & & \nonumber\\
	&=& - 2 \left[ b_{h}(z-h)\psi(\tau, z-h) + b_{h}(z)\psi(\tau, z+h)\right] \nonumber\\
	&  & +  \Lambda(\tau) z^2 \psi(\tau, z) \:.
	\label{eq:Schr3}
\end{eqnarray}
Let us set $b_h (z) = 0$ for all $z \le -1-h$ and $z \ge 1$, as well as $\psi (t,z) = 0$ all $z \le -1-h$ and $z \ge 1+h$. 
It is worthwhile to note that if this equation is fulfilled for $z \in [-1-h,1+h]$ (it will be trivially fulfilled outside
this interval) then it is obviously also fulfilled for all $z_m$ ($m=-N/2,...N/2$) in the discrete description; therefore, 
Eq.~\eqref{eq:Schr3} is equivalent -- in the continuum approximation -- to Eqs.~\eqref{eq:SchrodingerEquationBoseHubbardAngular} 
and \eqref{eq:numerical}.

Equation~\eqref{eq:Schr3} will be one of our starting points in the following for constructing the control
schemes using the eSTA formalism. This equation is often rewritten further by recalling that for a
differentiable function $f(x)$ it holds that $f(x \pm \epsilon) = e^{\pm \epsilon\partial_{x}}f(x)$. 
As a result, we arrive at the equation $i h\partial_\tau \psi(\tau, z) = H_2 \psi(\tau,z)$, where
\begin{eqnarray}
H_2 &=& - 2 \left[ e^{-ip} b_{h}(z) + b_{h}(z)e^{ip} \right] \nonumber\\
& &+ \Lambda(\tau) z^2 \label{ham_2}
\end{eqnarray}
and $p = -ih\partial_z$.

We now want to derive an approximated version of Eq.~\eqref{ham_2}.
By performing a Taylor expansion of both the $ e^{\pm i p} $ part and
the function $b_{h}(z)$ up to the second order in $h$, we obtain
\begin{eqnarray}
e^{\pm i p} &=& e^{\pm h \partial_z}  \approx 1 \pm h 
\partial_{z} - \frac{1}{2}h^2\partial^2_{z} \:, \label{eq:TaylorExpansionExp }                         \\
b_{h}(z) & \approx &  1 + h \partial_{h}b_{h}(z)|_{h = 0} + 
\frac{1}{2}h^2\partial_h^2b_{h}(z)|_{h = 0} \label{eq:TaylorExpansionB} \:.
\end{eqnarray}
For a small $h$, neglecting a constant energy shift, we get approximately
the Hamiltonian~\cite{Shchesnovich+Trippenbach:08}
\begin{eqnarray}\label{ham_1}
H_1 = -h^2\partial_z b_{0}(z) \partial_z +\Lambda(\tau) z^2 - 2b_{0}(z) \:,
\end{eqnarray}
where $b_{0}(z) = \sqrt{1-z^2}$. In the following, the Hamiltonian in Eq.~\eqref{ham_1} 
will serve as a second, alternative starting point for constructing the control schemes 
using the eSTA formalism.

If we assume, in addition, that $z$ is small (i.e. the population difference between 
the two states is small compared to the total particle number $N$) and neglect a 
constant energy shift, we obtain the Hamiltonian of a harmonic oscillator
\begin{equation}\label{ham_ho}
H_{0} = -h^2 \partial^2_{z} + \left[\Lambda(\tau) + 1\right] z^2 \:.
\end{equation}
Note that this last Hamiltonian has been the starting point for deriving the STA scheme in Ref.~\cite{JuliaDiaz+:12}.
In the following, we will use it as our primary point of departure for developing an eSTA-type 
control scheme.

The control task that we will address in the following now reduces to finding a control function $\Lambda(t)$ such
that we start at $\tau=0$ in the ground state of the Hamiltonian \eqref{eq:HamiltonianJosephsonJunctionSTA} with a
given $\Lambda(0)=\Lambda_i$ [resp. $U(0)=\hbar \Omega \Lambda_i/N=U_0$] and we end at $\tau=\tau_f$
in the ground state of the Hamiltonian \eqref{eq:HamiltonianJosephsonJunctionSTA} with a different given
$\Lambda(\tau_f)=\Lambda_f$ [resp. $U(\tau_f)=\hbar \Omega \Lambda_f/N=U_f$].

\subsection{Squeezing parameters} \label{SpinSqueezingParams}
Given that in the remainder of this work we will be concerned with spin-squeezed states, it is pertinent to
introduce at this point the relevant figures of merit of spin squeezing. In particular, two such quantities 
are the number-squeezing- and coherent spin-squeezing parameters. We define them in what follows
adopting the standard convention according to which the mean collective spin points in the direction of $J_x$ and 
the direction of minimal variance is that of $J_z$ [cf. Eq.~\eqref{PseudoAngular}]. The length of the collective 
spin will be denoted by $J\equiv N/2$.

The number-squeezing parameter, which in the problem at hand depends on time, is defined in terms 
of expectation values of the collective spin operators as~\cite{Wineland+:92}
\begin{equation}\label{NumberSqueezing}
\xi^2_{N}(t) = \frac{2}{J}\:\Delta J^2_z=
\frac{\Delta J^2_z}{N/4} \:,
\end{equation}
where $\Delta J^2_z \equiv \langle J^2_z \rangle - \langle J_z \rangle^2$ is the variance of
the operator $J_z$. Here $J/2 = N/4$ corresponds to the shot-noise limit, i.e. to the coherent spin state 
with $\langle J_z \rangle=0$. Therefore, a many-body state is number-squeezed ($\xi_{N}<1$) if 
the variance of one spin component is smaller than the shot-noise limit.

The (time-dependent) coherent spin-squeezing parameter is defined as~\cite{Sorensen+:01}
\begin{equation}\label{CoherentSpinSqueezing}
\xi^2_{S}(t) = \frac{N\Delta J^2_z}{\langle
J_x \rangle^2} = \frac{\xi^2_{N}(t)}{\alpha^2(t)} \:,
\end{equation}
where $\alpha(t)\equiv\langle \Psi(t)|2J_x/N|\Psi(t)\rangle$ is the measure of the phase coherence of the 
many-body state $|\Psi(t)\rangle$. The parameter $\xi_{S}$ quantifies the complex interplay between an improvement 
in number squeezing and loss of coherence. Importantly, this parameter can be used to quantify precision 
gain in interferometry; namely, the interferometric precision is increased to $\Delta\theta=\xi_{S}/\sqrt{N}$ (compared 
to the standard quantum limit $1/\sqrt{N}$) for spin-squeezed states~\cite{Gross:12}. Moreover, the inequality $\xi^2_{S}<1$ 
signifies that the many-body state in question is entangled~\cite{Sorensen+:01}.

\section{Application of eSTA formalism} \label{ApplicationESTA}
The general formalism of eSTA was derived in \cite{Whitty+:20, Whitty+:22, Whitty2022b}.
Its starting point is the exact system Hamiltonian $H_S$. This Hamiltonian can be approximated by
a Hamiltonian $H_0$, where a control function $\Lambda_0(t)$ can be derived for $H_0$ using STA techniques;
this control function $\Lambda_0(t)$ results in a fidelity equal to one for $H_0$.
However, when this control scheme is applied to the exact system Hamiltonian $H_S$, the fidelity will in general
obviously be lower than one. However, by making use of the eSTA formalism, we can then calculate analytically
an improved control scheme $\Lambda(t) = \Lambda_0 (t) + P_{\vec\lambda}(t)$ based on the knowledge of the
STA control scheme for $H_0$ that results in a larger fidelity when applied to the exact system Hamiltonian
$H_S$ (see below for the definition of $P_{\vec\lambda}(t)$ and more details).

In what follows, we apply the general eSTA framework to the preparation of spin-squeezed states
in internal BJJs. We will consider two different approaches in the following.
In the first, simplified approach, we assume that the system Hamiltonian is given by the approximated
Hamiltonian $H_{S,1}=H_{1}$ given by Eq.~\eqref{ham_1}.
This Hamiltonian can be approximated by $H_{0}$ given by Eq. \eqref{ham_ho} and an STA control scheme
$\Lambda_0$ has been derived for this Hamiltonian in Ref.~\cite{JuliaDiaz+:12}. This will be the first 
starting point for applying the general formalism of eSTA in the problem at hand.

In the second approach, we assume that the system Hamiltonian is given by the exact Hamiltonian $H_{S,2}=H_{2}$ 
of Eq.~\eqref{ham_2}, which is equivalent to $H_{\textrm{IBJJ}}$. This Hamiltonian can again be approximated by 
$H_{0}$ given by Eq.~\eqref{ham_ho}. This will be the second starting point for applying the general formalism 
of eSTA in this work.

In the problem under consideration, the approximated Hamiltonian $H_0$ is given 
by Eq.~\eqref{ham_ho}. In Sec.~\ref{ReviewESTA} below, we first briefly review how an STA control 
scheme, characterized by the control function $\Lambda_0 (\tau)$, was designed in Ref.~\cite{JuliaDiaz+:12}
for the same Hamiltonian. Their resulting control function $\Lambda_0 (\tau)$ will be the basis for 
developing the enhanced STA scheme in the present work.

\subsection{Review of STA scheme for harmonic approximation $H_{0}$} \label{ReviewESTA}
The case of a time-dependent $U(t)$ and a constant $\hbar\Omega$ based on the harmonic approximation
was investigated in Ref.~\cite{JuliaDiaz+:12}, whose main result are summarized in what follows.

Derived using Lewis-Riesenfeld invariants, we show that the following wave function fulfills the time-dependent
Schr\"odinger equation for the harmonic oscillator Hamiltonian $H_{0}$ [see Eq.~\eqref{ham_ho}]
\begin{eqnarray}
\lefteqn{\chi_n(t,z) =} & & \nonumber\\
&&	\left(\frac{1}{ \pi z_0^2}\right)^{1/4} % Normalisation factor
\frac{1}{\sqrt{2^{n}! b(t)^2}}\: {\displaystyle e^{ i \varphi(t) - 
\frac{z^2}{2 z_0^2b(t)^2}+i\frac{b'(t) \tau}{4h b(t)}}} \nonumber\\
&\times&	H_{n}\left[\left(\frac{z}{z_0 b(t)}\right)\right] \:, % Hermite polynomial term
	\label{eq:chi}
\end{eqnarray}
where the auxiliary function $b(t)$ must be a solution of the Ermakov equation
\begin{eqnarray*}
b''(t) - \frac{\omega_0^2}{b(t)^3} + b(t) \omega(t)^2 = 0 \:.
\end{eqnarray*}
We have here also $\varphi(t) = -\int_0^t ds\, \frac{h (1+2n)}{\tau z_0^2 b(t)^2}$,
$\omega(t)^2 = \frac{4 (1 + \Lambda (t))}{\tau^2}$ and $z_0 = \sqrt{2 h/(\tau \omega(0))}$ and
$\omega_0 = \omega(0)$.
We now use inverse engineering by first fixing a function $b(t)$ that satisfies the boundary conditions 
$b(0)=1$, $b(t_f)=\gamma$, $b'(0)=b'(t_f)=b''(0)=b''(t_f)=0$, where $\gamma=\sqrt{\omega(0)/\omega(t_f)}$.
In the following, we will choose a polynomial of degree $6$ that satisfies these conditions.
By inverting the above equation, we obtain an explicit expression for the required physical 
control function $\Lambda(t)$ resp. $U(t)$:
\begin{eqnarray}
\Lambda_0 (t) = \frac{U(t)N}{\hbar\Omega} = -1 + 
\frac{\Lambda(0)}{b(t)^4} - \frac {\tau^2 b''(t)}{4 b(t)} \:.
\end{eqnarray}

\subsection{eSTA correction of the control function}
As mentioned above, the key idea of enhanced STA is to modify the STA control function by taking into account that 
the approximated Hamiltonian $H_0$ -- which is amenable to STA treatment -- is different from the exact system 
Hamiltonian $H_S$. We assume that the modified control function is $\Lambda (t) = \Lambda_0 (t) + P_{\vec\lambda}(t)$. 
We consider here a polynomial $ P_{\vec{\lambda}}(t)$ of degree $\nu+2$ ($\nu$ being a positive integer) that would 
take some values $ \vec{\lambda}  = (\lambda_{0},..., \lambda_{\nu+1})$ in the interval for some $ t \in [t_0, t_{\nu+1}]$ 
with $ \lambda_{0} = \lambda_{\nu+1} = 0 $, and $ t_{0}  = t_0 , t_{\nu+1} = t_{f} $, i.e. $P_{\vec\lambda}(t_j) = \lambda_j$.
It is helpful to use the Lagrange interpolation that would take the form
\begin{eqnarray}
\label{eq:LagrangeInterpolationPolynomial}
P_{\vec{ \lambda }}(t) = \sum_{j=0}^{\nu+1}\lambda_{j}
\prod_{\substack{k=0\\k\neq j }}^{\nu+1}\frac{t-t_{k}}{t_{j} - t_{k}} \:.
\end{eqnarray}
The expression for $P_{\vec{ \lambda }}(t)$ can be simplified even further by demanding that 
$ \lambda_{0} = \lambda_{\nu+1} = 0 $:
\begin{eqnarray}
\label{eq:LagrangeInterpolationPolynomialSimplified}
P_{\vec{ \lambda }}(t) = \sum_{j=1}^{\nu}\lambda_{j}
\prod_{\substack{k=1\\k\neq j }}^{\nu}\frac{t-t_{k}}{t_{j} - t_{k}} \:.
\end{eqnarray}

These corrections $\vec{\lambda}$ are calculated by making approximation about the value of the fidelity 
landscape. We start by defining the auxiliary functions $G_n$, $K_{n}$ and $\mathrm{H}$ that will later 
be used to calculate $\vec{\lambda}$~\cite{Whitty+:20, Whitty2022b}.
Let $\Delta H = H_S - H_0$ be the difference between the exact system Hamiltonian and the approximated one.
Using the STA wave functions $\chi_{n}$ of the approximated Hamiltonian [cf. Eq. \eqref{eq:chi}] we can 
evaluate $ G_n $ using the formula
\begin{eqnarray}
G_{n} = \int_{0}^{t_{f}} dt \braket{\chi_n}{\Delta H |\chi_{0}} \:. \label{eq:Gn}
\end{eqnarray}
Similarly, we can calculate $\vec{K}_{n} $ as
\begin{eqnarray}
\vec{K}_{n} = \int_{0}^{t_{f}} dt \braket{\chi_{m}}{\vec{\nabla} H_S|\chi_{0}} \:, \label{eq:Kn}
\end{eqnarray}
where $\vec{\nabla} H_S$ is the gradient of the Hamiltonian with respect to the control parameter $\vec\lambda$.
In addition, let us define the Hessian matrix $\mathcal{H}$ with matrix elements
\begin{eqnarray}
\mathcal{H}_{l,k} = \sum_{n = 1}^{\mathcal{N}}\left[G_{n}(W_{n})_{l,k} + 
\left(  \vec{K}^{*}_{n}\right)_{k}\left(  \vec{K}_{n} \right)_{l}\right] \:,\label{eq:HessianMatrix}
\end{eqnarray}
where $ (W_{n})_{l,k} = \braket{\chi_n}{\partial_{\lambda_l}\partial_{\lambda_k} H_S | \chi_0}$ is a
matrix evaluated by taking the second derivative with respect to the control parameter and
$\mathcal{N}$ is the number of STA wave functions that we take into account.

We can now calculate the correction parameters $\vec\lambda$ using $G_n$ and $K_n$, assuming
that a fidelity equal to unity can be achieved for the exact Hamiltonian~\cite{Whitty+:20}.
Alternatively, the correction parameters $\vec\lambda$ can be computed using $G_n$, $K_n$, and
the values of $\mathcal{H}_{l,k}$~\cite{Whitty2022b}. We will follow the second procedure and the correction
parameters $\vec\lambda$ can be calculated via
\begin{eqnarray}
	\vec\lambda = \frac
	{\vec{v}\norm{v}^{2}}
	{\vec{v}^{T}\mathrm{H}\vec{v}} \:,
	\label{eq:eSTAcorrectionsHessian}
	\label{esta_labda}
\end{eqnarray}
where
\begin{eqnarray}
	\vec{v} = \sum_{n=1}^{\mathcal{N}}\text{Re}(G_{n}^{*}\vec{K}_{n})\:.
	\label{eq:FidelityVector}
\end{eqnarray}

In the first eSTA-based approach, we choose $H_{S,1}=H_{1}$, as already mentioned above. In this case we have
\begin{equation}
\Delta H_1 = -h^2 \partial_z  b_{0}(z)
\partial_z + h^2 \partial^2_z - 2 b_{0}(z) - z^2 \:,
\end{equation}
where we can see that the control function $\Lambda (t)$ is cancelled out here.
Recalling the form of $H_{1} $ [cf. Eq.~\eqref{ham_1}], we can see that the only part that depends on $ \lambda_{i} $ 
is the $ z^2 $ term. Now taking the gradient of $ H_{1}$ with respect to the control parameters in this case only 
amounts to performing the following derivatives
\begin{eqnarray}
\lefteqn{\partial_{\lambda_i} H_{S,1} = \partial_{\lambda_i} H_{1} = \partial_{\lambda_i} \tilde{ \Lambda }(\tau) z^2} & & \nonumber\\
	&=& \partial_{\lambda_i} \left[ \Lambda_0 z^2 + P_{\vec{ \lambda }}(t)z^2 \right]
	= z^2 \prod_{\substack{k=2\\k\neq i }}^{n-1}\frac{t-t_{k}}{t_{i} - t_{k}} \:.
	\label{eq:GradientLambda}
\end{eqnarray}
This results in
\begin{eqnarray}
	\label{eq:KnOverSpace}
	\Braket{\chi_{m}|\partial_{\lambda_i} H_{S_1}|\chi_{0}} = \prod_{\substack{k=2\\k\neq i }}^{n-1}
	\frac{t-t_{k}}{t_{i} - t_{k}} \Braket{\chi_{m}|z^2 |\chi_{0}} \:.
\end{eqnarray}
Due to the symmetry and properties of $\chi_n$, it follows that this expression has a nonzero value 
only for $m=2$ and, therefore, only $\vec K_2$ is nonzero in this case. Because $\Lambda$ is only linear 
in $\vec\lambda$, it also follows that $W_{n} = 0$.
In this way Eqs.~\eqref{eq:FidelityVector} and \eqref{eq:HessianMatrix} simplify significantly, and we can write
\begin{align}
& \vec{v} = \text{Re}(G_{2}^{*}\vec{K}_{2}) \label{eq:FidelityVectorSimple} \\
& \mathcal{H}_{l,k} = \left(  \vec{K}^{*}_{2} \right)_{k}\left(  
\vec{K}_{2} \right)_{l} \:. \label{eq:HessianMatrixSimple}
\end{align}
The correction parameters $\vec\lambda_1$ in this approach are then given by Eq. \eqref{esta_labda}. 
In the following, we set $\nu=5$ and denote the resulting control function by $\Lambda_1$.

In the second eSTA-based approach, we choose $H_{S,2}=H_{2}$ as already mentioned above. In this case we have
\begin{equation}
	\Delta H_2 =
	- 2 \left[ e^{-ip} b_{h}(z) + b_{h}(z)e^{ip} \right] + h^2 \partial^2_{z} - z^2 \:,
\end{equation}
where we can see that the control function $\Lambda (t)$ is again cancelled out here.
The form of $H_{2}$ [cf. Eq.~\eqref{ham_2}] implies that again the only part dependent on $ \lambda_{i}$
is the $ z^2 $ term. Therefore, similar to the above case we obtain
\begin{eqnarray}
	\label{eq:KnOverSpace2}
	\Braket{\chi_{m}|\partial_{\lambda_i} H_{S_2}|\chi_{0}} = \prod_{\substack{k=2\\k\neq i }}^{n-1}\frac{t-t_{k}}{t_{i}
	- t_{k}} \Braket{\chi_{m}|z^2 |\chi_{0}} \:,
\end{eqnarray}
which is nonzero only for $m=2$, resulting in similar expressions as Eq.~\eqref{eq:HessianMatrixSimple}, however
with a different value of $G_2$. The corresponding correction parameters $\vec\lambda_2$ in the second approach are
then again given by Eq.~\eqref{esta_labda}. In the following, we consider the cases of $\nu=5$ (the resulting 
control function will be denoted $\Lambda_2$) and $\nu=1$ (the control function denoted by $\widetilde\Lambda_2$).
In the following section we will examine these three resulting control schemes based on the eSTA formalism.

% --------------------------------------------------------------------------------------------------
% Results and Discussion
% --------------------------------------------------------------------------------------------------

\section{Results and discussion\label{sect_results}}
In this section, we apply the eSTA scheme -- as described in the previous section -- to a BJJ.
We apply the eSTA protocol to both the Hamiltonian $H_{1}$ and $H_{2}$ with $\nu = 5$ corrections (control functions
$\Lambda_1$ and $\Lambda_2$), as well as applying the protocol to $H_{2}$ with $\nu = 1$ corrections (control
function $\widetilde\Lambda_2$).

We will compare the results of different eSTA schemes between each other and with the results obtained using
the original STA scheme $\Lambda_0$, as well as with the results of an adiabatic control scheme $\Lambda_{ad}(t)$. 
The adiabatic scheme is a polynomial of degree $3$ determined by the conditions $\Lambda_{ad}(0)= \Lambda_i$, 
$\Lambda_{ad} (t_f)= \Lambda_f$, $\Lambda'_{ad} (0)= \Lambda'_{ad}(t_f)=0$. For better comparison with previous 
works, we introduce the Rabi time $t_R= 2\pi /\Omega$ and use it as the characteristic timescale in the problem
at hand; in other words, the squeezed-state preparation times in the following will be expressed in units of $t_R$.

Given that we have adopted the Rabi time $t_R$ as the characteristic timescale in the problem under 
consideration, it is of interest to consider relevant values of $t_R$ in the experimental realizations of internal 
BJJs. It is pertinent to do so using as a guide two of the first experiments demonstrating spin squeezing in these 
systems~\cite{Gross+:10,Riedel+:10}. In particular, a Rabi coupling with the parameter $\Omega$ in the range from 
$0$ to $2\pi \times 600$ Hz was implemented in \cite{Gross+:10}, while $\Omega = 2\pi \times 10$ Hz was used in 
\cite{Riedel+:10}. Assuming values of $\Omega$ in the same range as in these experiments, for $\Omega = 2 \pi \times 
2$ Hz we obtain the Rabi time $t_R = 500$ ms, while for $\Omega = 2 \pi \times 10$ Hz we find $t_R = 2\pi/\Omega = 100$ ms.

\subsection{Target-state fidelity in different control schemes}\label{sect_fidelity}
We will first discuss the target-state fidelity defined as $F = \left| \langle \Psi_{T} | \Psi(t_f) \rangle \right|^2$,
where $\Psi_{T}$ is the target state, i.e. the ground state of the Hamiltonian $H_{\textrm{IBJJ}}$ in Eq.~\eqref{eq:HamiltonianJosephsonJunctionSTA}
for $t = t_f$, while $\ket{\Psi(t_f)}$ is the state obtained by the time evolution of the system driven by the Hamiltonian
$H_{\textrm{IBJJ}}$ \eqref{eq:HamiltonianJosephsonJunctionSTA} with the control scheme $\Lambda(t)$ applied.
We will consider the special case with $\Lambda(0)=0$ and $\Lambda(t_f) = 50$ in the following.

\begin{figure}
	\begin{center}
	\includegraphics[width=0.8\linewidth]{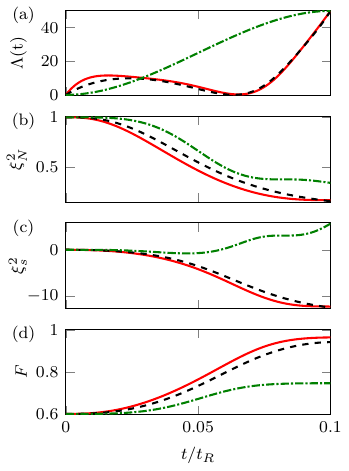}
	\end{center}
	\caption{Time evolution of (a) control function $\Lambda(t)$, (b) number squeezing $\xi_N^2(t)$,
		(c) coherent spin squeezing $\xi_S^2 (t)$ in dB, (d) fidelity $F(t)$.		
		Enhanced STA applied to discrete Hamiltonian (control function $\Lambda_2$, $\nu=5$: red, solid lines),
		STA (control function $\Lambda_0$, black, dashed lines), reference
		adiabatic scheme (control function $\Lambda_{ad}$, green, dashed-dotted lines).
		$N=10$, $t_f=0.1 t_R$, $\Lambda(0)=0$, $\Lambda(t_f) = 50$.
		\label{fig:example}}
\end{figure}

An example of this can be found in Fig.~\ref{fig:example}, where the time evolution of different features can be seen.
In this case we considered a system with $N=10$ particles and the evolution runs from $t=0$ to $t_f = 0.1\, t_R$.
In Fig.~\ref{fig:example}(a) we can see the different control functions $\Lambda(t)$ as function of time for the different
approaches, respectively adiabatic $\Lambda_{ad}$, STA scheme $\Lambda_0$ and the eSTA scheme $\Lambda_2$ (applied to $H_{2}$ 
with $\nu = 5$).

\begin{figure}[t]
	\begin{center}
	\includegraphics[width=0.8\linewidth]{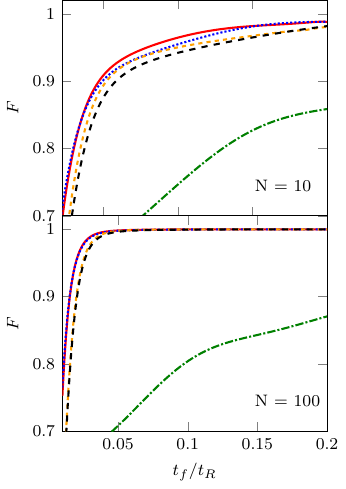}
	\end{center}
	\caption{Fidelities $F$ versus final  time $t_f$ for eSTA applied to Hamiltonian $H_2$ ($\Lambda_2$, $\nu=5$:
red, solid lines; $\widetilde\Lambda_2$, $\nu=1$: blue, dotted lines), eSTA applied to approximated 
Hamiltonian $H_1$ ($\Lambda_1$, orange,small dashed line), STA scheme ($\Lambda_0$, black, dashed lines) 
and as a reference adiabatic scheme ($\Lambda_{ad}$, green, dashed-dotted lines) with different particle numbers 
$N=10$ (above), $N=100$ (below). $\Lambda(0)=0$, $\Lambda(t_f) = 50$.
		\label{fig:fidelity}}
\end{figure}

The remaining figures are the time evolution of the number squeezing $\xi_N^2$ of Eq.~\eqref{NumberSqueezing} [Fig.~\ref{fig:example}(b)]
and the coherent spin-squeezing parameter $\xi_{S}^2$ of Eq.~\eqref{CoherentSpinSqueezing} expressed in dB [Fig.~\ref{fig:example}(c)],
as well as the fidelity $F$ [Fig. \ref{fig:example}(d)] when the three different control functions are applied.
We can already see that the eSTA scheme $\Lambda_2$ gives rise to a higher fidelity than the STA scheme $\Lambda_0$, without compromising
the achievable squeezing. To examine this in more detail, we plot the fidelity for different final times $t_f$ and for different eSTA
protocols ($\Lambda_1, \Lambda_2, \widetilde\Lambda_2$). The results are summarized in Fig.~\ref{fig:fidelity},
for $N=10$ and $N=100$ particles. We can clearly see that eSTA outperforms its STA counterpart even when only one correction
(control function $\widetilde\Lambda_2$) is used, or when it is applied to an approximated version of the original Hamiltonian
of the system (control function $\Lambda_1$). The effect is more pronounced for smaller particle numbers as the approximation
of $H_{\textrm{IBJJ}}$ with $H_{0}$ tends to lose its validity, thus increasing the effects inherent to the eSTA approach.

%
%
% ------------------------ Fig. 2 -------------------------------
% ---------------------------------------------------------------
%
%

%
%
% ------------------------ Fig. 3 -------------------------------
% ---------------------------------------------------------------
%
%

%
%
% ------------------------ Fig. 4 -------------------------------
\begin{figure*}[t]
	\begin{center}
	\includegraphics[width = \linewidth]{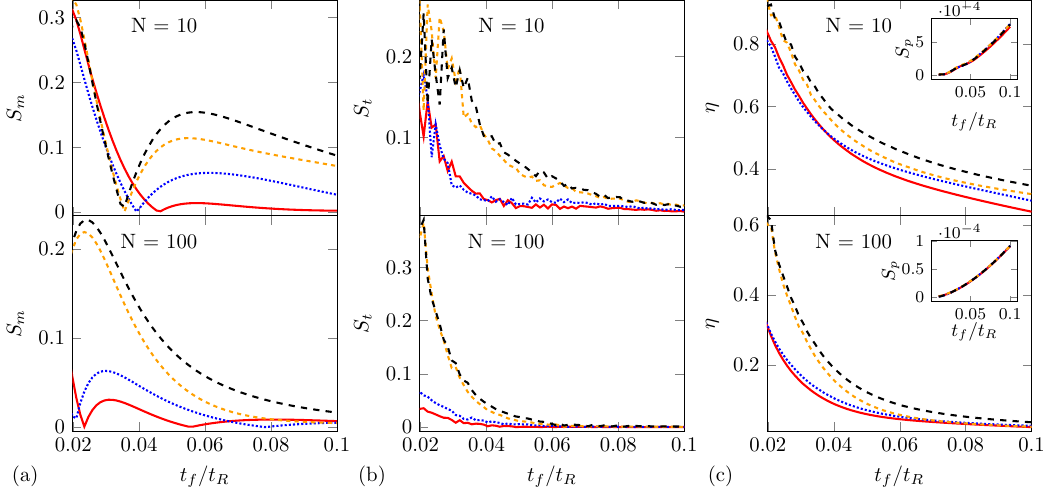}
	\end{center}
	\caption{
(a) Sensitivity $S_m$ for systematic amplitude error, (b)  sensitivity $S_t$ for systematic time-shift, 
(c) imperfection $\eta$; inset: sensitivity $S_p$ for phase noise.
eSTA applied to Hamiltonian $H_2$ ($\Lambda_2$, $\nu=5$:
red, solid lines; $\widetilde\Lambda_2$, $\nu=1$: blue, dotted lines), eSTA applied to approximated Hamiltonian $H_1$ ($\Lambda_1$, 
orange, small dashed line), STA scheme ($\Lambda_0$, black, dashed lines).Different particle numbers: $N=10$ (above), $N=100$ (below).
The best result is obtained for $\Lambda_2$,  $\nu = 5 $ corrections and eSTA applied to the original Hamiltonian (solid red line) 
as it ensures lowest sensitivity to systematic errors and highest fidelity.
		\label{fig:robustness}}
\end{figure*}
% ---------------------------------------------------------------
%
%

\subsection{Stability of the control schemes} \label{StabilityControlScheme}
An important property of the control schemes is their robustness. Therefore, in the following, we will 
consider systematic errors, i.e. an unknown, constant error in the experimental setup. First, we consider 
a systematic error in the amplitude of the control function $\Lambda(t)$ of the form $\Lambda_{\delta,m}= 
(1 + \delta) \Lambda(t)$ for $ t \in [0, t_f]$ and for an unknown constant value of $\delta$.
We calculate numerically the systematic error sensitivity of the control scheme:
\begin{eqnarray}
S_m = \left|\frac{\partial F}{\partial \delta}\right|_{\delta=0} \:.
\label{eq_S}
\end{eqnarray}
This number is a measure of the sensitivity of the control scheme against systematic errors in the
amplitude of the control function and the lower the number, the more stable the protocol is.
The result can be seen in Fig.~\ref{fig:robustness}(a).
It shows how the eSTA protocol applied to the full Hamiltonian with $\nu = 5$
corrections is the most robust against systematic error, but a considerably lower sensitivity
is already achieved for $ \nu = 1$ correction when compared to the STA protocol.

The second case is the systematic error in the time of the control function $\Lambda(t)$ of the form
$\Lambda_{\delta,t}= \Lambda(t+\delta)$ for $ t \in [-\delta, t_f - \delta]$ and $\Lambda_{\delta,t} =
	\Lambda(t)$ for $t \notin [-\delta, t_f - \delta]$ for an unknown constant value of $\delta$.
The sensitivity of the control scheme against this systematic error $S_t$ is defined similar to Eq.~\eqref{eq_S}.
The results in Fig.~\ref{fig:robustness}(b) confirms once again what we have seen for both $F$ and $S_m$.
In this case we would like to point out how the line relative to the eSTA applied to $H_{1}$ ($\Lambda_1$)
is closer to the STA line ($\Lambda_0$) when compared to the ones obtained applying eSTA to $H_{2}$ ($\Lambda_2$
and $\widetilde\Lambda_2$).
This can be explained by the fact that the approximation of $H_{1}$ with $H_{0}$ is valid so the effects of
the eSTA approach are less pronounced when compared to the STA one.

It is pertinent to consider at this point the effects of environmental noise. As an example, we 
will consider (classical) phase noise being coupled to $J_z$~\cite{Ferrini2010}. To be more specific, 
we assume the noise to be of the form $\xi(t) \hbar\Omega J_z$, where $\xi(t)$ describes 
a stochastic process with (classical) Gaussian white noise whose strength is quantified by the parameter $\mu$. 
The corresponding sensitivity can be defined in the form of the dimensionless quantity
\begin{eqnarray}
S_p &=& t_R \fabs{\frac{\partial F}{\partial \left( \mu^2 \right)}}\, .
\end{eqnarray}
[Note that the dimensionless character of $S_p$ is a consequence of the fact that $\mu^2$ 
has the dimensions of time.] As follows from a general framework presented in Ref.~\cite{Whitty+:22},
this sensitivity can be evaluated as
\begin{eqnarray}\label{eq:S_p}
S_p &=& t_R \Omega^2 
\Bigg\vert
\int_0^t ds 
\nonumber \\
&\times& \Big[\text{Re} \left\{ 
\braXket{\Psi_T(s)}{ J_z^2}{\Psi_0(s)}\braket{\Psi_0(s)}{\Psi_T(s)}
\right\}
\nonumber \\
&&-
\fabsq{\braXket{\Psi_T(s)}{J_z}{\Psi_0(s)}}
\Big]
\Bigg\vert \:,
\end{eqnarray}
where $\ket{\Psi_0 (0)}$ is the initial state of the system (coherent spin state in the problem at hand),
$\ket{\Psi_T (t_f)}$ its target state (spin-squeezed state in the problem under consideration), while
$\ket{\Psi_0 (t)}$ is the time-evolved solution of the Schr\"odinger equation at time $t$. What makes the 
quantity $S_p$ in Eq.~\eqref{eq:S_p} particularly useful is the fact that it allows one to characterize 
the sensitivity of the system to environmental noise without having to numerically simulate the full 
open-system dynamics.

The obtained results for the sensitivity $S_p$ are illustrated in the inset of Fig.~\ref{fig:robustness}(c). 
What can be inferred from those results is that the sensitivity $S_p$ is very similar for the STA and eSTA schemes; $S_p$ 
is very small (compared to $S_m$ and $S_t$) and increases with increasing total control (state-preparation) time $t_f$.

We also incorporate $S_p$ in an additional figure of merit that encapsulates both the fidelity and 
the sensitivities to systematic errors defined above. This quantity, which in the following will be referred to 
as imperfection, is defined as
\begin{equation}\label{eq:imperfection}
\eta = \sqrt{(1-F)^{2} + S_m^{2} + S_t^2 + S_p^2} \:,
\end{equation}
where $F$ is the fidelity and $S_m, S_t, S_p$ are the sensitivities. A small value of $\eta$
corresponds to low infidelity (i.e. high fidelity) and small sensitivities (i.e. high degree of robustness) to 
systematic errors and phase noise. Therefore, the lower the value of $\eta$, the better the control scheme.
The results shown in Fig.~\ref{fig:robustness}(c) summarize findings in the remainder 
of this paper. These results indicate that the best performance is achieved using the eSTA method with 
higher number of corrections, applying this method to the original (approximation-free) Hamiltonian of 
the system.

\subsection{Comparison with other control schemes for generating spin squeezing in internal BJJs}
In this section we compare the amount of squeezing obtained in internal BJJs using the eSTA protocol proposed here with
the ones obtained with previously proposed control schemes. In general, for different final times $ t_f $, the squeezing
obtained using the eSTA protocol is similar to the one obtained with the STA protocol. What makes the eSTA protocol more
powerful than its STA counterpart is the fact that it is much more robust against systematic errors. To illustrate that,
let us consider the example for $ t_f/t_R = 0.05$ with $ N = 100 $ particles and $ \Lambda(t_f) = 50 $. We will focus only
on the eSTA protocol applied to the full Hamiltonian with $\nu = 5$ corrections, as it has already been shown that this is
the best performing one among possible eSTA-type protocols.

To make contact with anticipated experimental realizations of our proposed squeezed-state preparation 
scheme, it is pertinent to consider typical state-preparation times. For example, with a value of $t_R = 500$ ms 
($\Omega = 2 \pi \times 2$ Hz) for the Rabi time the corresponding state-preparation time is $t_f= 0.05\:t_R = 25$ ms.
Alternatively, for $t_R = 2\pi/\Omega = 100$\:ms ($\Omega = 2 \pi \times 10$ Hz) we obtain $t_f= 0.05\:t_R = 5$ ms.
Both of these times are in the same range as the typical times needed for the generation of spin squeezing in existing  
experiments; for instance, this time was around $20$ ms in the experiment of Ref.~\cite{Gross+:10} and approximately 
$15$ ms in that of Ref.~\cite{Riedel+:10}.

In the experiment of Ref.~\cite{Gross+:10} the nonlinear interaction parameter had the value $U/\hbar = 
2\pi \times 0.063$\:Hz, which along with $\Omega=2\pi\times 2$\:Hz results in a ratio $U/(\hbar\Omega)=0.03$. We here 
consider on purpose a significant variation of the nonlinear interaction parameter, with the corresponding change of 
the control parameter $\Lambda(t)$ from $\Lambda(0)/N=U(0)/(\hbar\Omega)=0$ to $\Lambda(t_f)/N=U(t_f)/(\hbar\Omega)=0.5$ 
($N=100$), to demonstrate that our scheme for the generation of spin squeezing works even far away from the adiabatic 
regime. As already stated above, the total squeezed-state preparation time here is then $t_f= 0.05\, t_R = 25$\:ms,
which is quite comparable to the state-preparation time of $20$\:ms in Ref.~\cite{Gross+:10}.

For this case of $ t_f/t_R = 0.05$, $N = 100$ particles, and $ \Lambda(t_f)/N = 0.5$,
the squeezing obtained by means of the eSTA protocol is $ \xi_S^2 = - 16.9$\:dB, while the one obtained
with the STA protocol is $ \xi_S^2 = - 15.8$\:dB. Hence, the two protocols are comparable in terms of achievable
spin squeezing. However, the eSTA protocol is more robust against systematic errors, as the value of the sensitivity
$ S_m $ is $ 0.005 $ for the eSTA protocol and $0.086$ for its STA counterpart; this is an improvement by a factor
of around $17$. Similar results are obtained for the systematic error in the duration of the control scheme, where 
the sensitivity $S_t $ is $ 0.0012 $ for the eSTA protocol and $ 0.017 $ for the STA one; this amounts to an 
improvement by a factor of around $14$.

The behaviour is more prominent for shorter final times and it is not strongly affected by the particle number.
The fact that it does not compromise the squeezing achievable using the STA protocol, but is -- at the same time -- much
more robust against systematic errors, makes the eSTA protocol proposed here an excellent candidate for the experimental
realization of spin-squeezed states in internal BJJs.

For the sake of completeness, it is worthwhile noting that the generation of spin-squeezed states in 
internal BJJs, governed by the Lipkin-Meshkov-Glick-type Hamiltonian of Eq.~\eqref{eq:HamiltonianJosephsonJunctionSTA}, 
belongs to the twist-and-turn dynamical scenario for creating spin squeezing~\cite{Muessel+:15,Sorelli+:19}. This type
of dynamics, in which the conventional spin-squeezing dynamics engendered by the (nonlinear) one-axis-twisting term $J_z^2$
is altered by the simultaneous presence of the (linear) turning term $J_x$, was experimentally investigated~\cite{Muessel+:15} 
under the same assumption used in the present work -- namely, that the initial state of the system is a spin coherent state 
pointing in the $x$ direction. The twist-and-turn dynamics allows the preparation of highly entangled, metrologically
relevant states (i.e. cat-like states), with preparation times that are logarithmic in the system size (i.e. the size of the collective 
spin), starting from this last, uncorrelated spin coherent state; this is a feature that the twist-and-turn dynamics share with that 
of two-axis countertwisting~\cite{Kitagawa+Ueda:93}. Moreover, an investigation of the twist-and-turn dynamics in the short-time 
regime has already demonstrated that those dynamics are optimal for the generation of spin squeezing~\cite{Sorelli+:19}, 
being faster than in the one-axis-twisting case.

The existing experimental demonstration of the twist-and-turn spin-squeezing dynamics in internal BJJs 
is based on the idea of abruptly switching the nonlinear interaction to a finite value -- i.e. performing a quench of 
nonlinear coupling -- in the presence of fixed linear Rabi-type coupling~\cite{Muessel+:15}. While this established 
experimental approach utilized a Feshbach resonance for increasing the nonlinear coupling strength, for our envisioned scheme -- in 
which this coupling strength is changed in time in a rather smooth fashion [as illustrated in Fig.~\ref{fig:example}(a)] -- the 
approach of Ref.~\cite{Riedel+:10} could even be more suitable. In that approach, which entails spatially inhomogeneous 
microwave fields, the trapping potentials of the atom cloud in the two relevant hyperfine states can be manipulated through 
microwave level shifts.

While the twist-and-turn dynamics with the fixed ratio of the nonlinear- and linear coupling strenghts 
has already been shown to be locally optimal for the generation of spin squeezing~\cite{Sorelli+:19}, at least in the absence 
of losses, it can be argued that our envisioned eSTA-based approach can yield comparable performance in anticipated realistic 
experimental realizations. Firstly, the eSTA formalism, which was inspired in part by optimal-control techniques~\cite{Whitty+:20},
has already been shown to yield results very close to the relevant quantum speed limits in certain classes of quantum-control 
problems (e.g. in the context of coherent atom transport~\cite{Hauck+Stojanovic:22,Whitty+:22}). Secondly, the extraordinary 
robustness of the eSTA-based control schemes to various experimental imperfections, compared to both its STA counterparts and 
other control schemes [cf. Sec.~\ref{StabilityControlScheme}], bodes well for experimental realizations; on the other hand, 
the robustness of optimal-control schemes is problem-specific and is not guaranteed to be consistently better than that of 
other types of control schemes (for an example of a state-preparation problem in a system with a similar underlying Hamiltonian, 
see, e.g., Refs.~\cite{Nauth+Stojanovic:22} and \cite{Stojanovic+Nauth:22}). Finally, the quantitative effect of atomic losses 
in our scheme -- especially that of two-body spin-relaxation losses for $F = 2$ hyperfine states -- is yet to be investigated; 
the combined effects of two- and three-body losses were, for example, shown to lead to a decrease of around $22\%$ in the 
achievable optimal spin squeezing in the experimental realization of Ref.~\cite{Muessel+:15}.

\section{Summary and conclusions} \label{SummConcl}
To summarize, in this paper we revisited the problem of generating strongly spin-squeezed states in internal bosonic
Josephson junctions with time-dependent interspecies interaction strength. Starting from the standard Lipkin-Meshkov-Glick 
type Hamiltonian of this system, we designed a robust state-preparation scheme using the recently proposed method of enhanced 
shortcuts to adiabaticity. We quantitatively characterized the quantum dynamics underlying the envisioned preparation of 
spin-squeezed states by computing the time dependence of the target-state fidelity, as well as that of the coherent 
spin-squeezing and number-squeezing parameters.

We demonstrated that our scheme for generating spin-squeezed states yields better results than the previously proposed
protocols based on shortcuts to adiabaticity. Importantly, we demonstrated that the inherent increased robustness of enhanced 
shortcuts to adiabaticity -- compared to their parent method -- makes our state-preparation scheme more amenable to experimental 
implementations than the previously proposed protocols for generating spin-squeezed states in bosonic Josephson junctions.
To further facilitate such experimental implementations, a future work could include an investigation of other sources of 
decoherence (aside from the phase noise already discussed in the present work) in this system~\cite{Sinatra+:12}, e.g. the 
effect of atomic losses on the achievable spin squeezing.\\

\begin{acknowledgments}
	V.M.S. acknowledges a useful discussion with P. Treutlein.
	M.O and A.R acknowledge that this publication has emanated from research
	supported in part by a research grant from Science Foundation Ireland (SFI)
	under Grant Number 19/FFP/6951.
	This research was also supported by the Deutsche Forschungsgemeinschaft
	(DFG) -- SFB 1119 -- 236615297 (V.M.S.).
\end{acknowledgments}

\end{document}